\documentclass[12pt]{iopart}
\usepackage{iopams}
\usepackage{bbold}
\usepackage{graphicx}
\usepackage{float}
\usepackage{setstack}

\begin{document}

\title[]{Quantum discord of $SU(2)$ invariant states}

\author[]{B. \c{C}akmak, Z. Gedik}
 \address{Faculty of Engineering and Natural Sciences, Sabanci University, Tuzla, Istanbul 34956, Turkey}
 \ead{cakmakb@sabanciuniv.edu}

\begin{abstract}
We have analytically calculated the quantum discord for a system composed of spin-$j$ and spin-$1/2$ subsystems possessing $SU(2)$ symmetry. We have compared our results with the quantum discord of states having similar symmetries and seen that in our case amount of quantum discord is much higher. Moreover, using the well known entanglement properties of these states, we have also compared their quantum discord with entanglement. Although the system under consideration is separable nearly all throughout of its parameter space as $j$ increases, we have seen that discord content remains significantly large. Investigation of quantum discord in $SU(2)$ invariant states may find application in quantum computation protocols that utilize quantum discord as a resource since they arise in many real physical systems.
\end{abstract}

\section{Introduction}

Multipartite quantum states contain different kinds of correlations which can or cannot be of classical origin. Entanglement has been recognized as the first indicator of non-classical correlations and lies at the heart of quantum information science \cite{1}. It has been considered as the resource of almost all protocols in quantum computing. However, recent research on quantum correlations has shown that entanglement is not the only kind of useful quantum correlation. Quantum discord (QD), which is defined as the discrepancy between two classically equal descriptions of quantum mutual information, has also proven to be utilizable in quantum computing protocols \cite{2,3}. Moreover, QD is more general than entanglement in the sense that it can be present in separable mixed quantum states as well. Following this discovery, much effort has been put into investigating the properties and behavior of QD in various systems ranging from quantum spin chains to open quantum systems \cite{4}. Nevertheless, since evaluation of QD requires a very complex optimization procedure, the significant part of the development in the field is numeric and analytical results are present only for some very restricted set of states. In general, these restrictions are introduced by forcing certain symmetries and limiting the size and the dimension of the system under consideration. A short list of analytical results would include the progress in, $X$-shaped states of different dimensions \cite{5,6,7,8,9}, $2\otimes d$ dimensional two-parameter class of states \cite{10}, $d\otimes d$ dimensional Werner and pseudo-pure states \cite{11}, general real density matrices displaying $Z_2$ symmetry \cite{12}, two-mode Gaussian states \cite{13}, and $2\otimes d$ dimensional mixed states of rank-$2$ \cite{14,15,16,17} where $d$ denotes the Hilbert space dimension of the system under consideration. QD witnesses have also been introduced for $2\otimes d$ systems \cite{18}. Following QD, many other quantum and total correlation quantifiers have been introduced \cite{19,20,21,22,23,24}.

Bipartite $SU(2)$ invariant states are defined by their invariance under rotation of both spins, $U_1\otimes U_2\rho U^{\dag}_1\otimes U^{\dag}_2=\rho $, where $U_{1(2)}=\exp(i\vec{\alpha}\cdot\vec{S}_{1(2)})$ is the usual rotation operator and the length of $\vec{\alpha}$ is chosen according to the spin length $|\vec{S}|$ \cite{25,26}. In other words, these states commute with every component of the total spin operator $\vec{J}=\vec{S}_1+\vec{S}_2$. In real physical systems, $SU(2)$ invariant density matrices arise when, for example, considering reduced state of two spins described by a $SU(2)$ invariant Hamiltonian or multi-photon states generated by parametric down-conversion and then undergo photon losses \cite{27}. Entanglement structure of states under certain symmetries has been vastly explored in the literature \cite{28,29,30}. For $SU(2)$ invariant states, which is central to this work, negativity has shown to be a necessary and sufficient criterion for separability \cite{25,26} and relative entropy of entanglement has been analytically calculated \cite{31} for $(2j+1)\otimes 2$ and $(2j+1)\otimes 3$ dimensional systems. Furthermore, entanglement of formation (EoF), a measure which also involves a complex optimization procedure, has been analytically evaluated in the case of mixed $(2j+1)\otimes 2$ dimensional systems \cite{32}. Recently, a very closely related article came up, which evaluates different correlation measures that are more general than entanglement, in $\cal{O}\otimes\cal{O}$ invariant states \cite{33}. However, to the best of our knowledge, this is the first attempt to explore the quantum discord in $(2j+1)\otimes 2$ dimensional states.

In this work, we have analytically calculated the QD of $SU(2)$ invariant $(2j+1)\otimes 2$ dimensional system. We have compared our results with the entanglement properties of these states and other analytical calculations of quantum discord in systems having similar symmetries. We have observed that while entanglement content decreases as $j$ increases, amount of QD remains significantly larger with its maximum value also following a decreasing trend.

\section{Quantum Discord}

In this section, we shall review the concept of quantum discord. We have very briefly mentioned that quantum discord is the difference between the quantum extensions of the classical mutual information. First and direct generalization of classical mutual information is obtained by replacing the Shannon entropy with its quantum analog, the von Neumann entropy
\begin{equation}
I(\rho^{ab})=S(\rho^a)+S(\rho^b)-S(\rho^{ab}).
\end{equation}
Here, $\rho^a$ and $\rho^b$ are the reduced density matrices of the subsystems and $S(\rho)=-\tr\rho\log_2\rho$ is the von Neumann entropy. On the other hand, in classical information theory, mutual information can also be written in terms of the conditional probability. However, generalization of conditional probability to quantum case is not straightforward since the uncertainty in a measurement performed by one party depends on the choice of measurement. Therefore, one has to optimize over the set of measurements made on a system \cite{2,3,4}
\begin{equation}
C(\rho^{ab})=S(\rho^a)-\underset{\{\Pi_k^b\}}{\min}\sum_kp_kS(\rho_k^a),
\end{equation}
where, in this work, \{$\Pi_k^b$\} is always understood to be the complete set of one-dimensional projective measurements performed on subsystem $b$ and $\rho_k^a=(I\otimes \Pi_k^b)\rho^{ab}(I\otimes \Pi_k^b)/p_k$ are the post-measurement states of subsystem $a$ after obtaining the outcome $k$ with probability $p_k=\tr(I^a\otimes\Pi_k^b \rho^{ab})$ from the measurements made on subsystem $b$. $C(\rho)$ can physically be interpreted as the maximum information gained about the subsystem $a$ after the measurements on subsystem $b$ while creating the least disturbance on the overall quantum system. This quantity is also referred as classical correlations contained in a state \cite{3,4}. Since classical versions of the aforementioned expressions for quantum mutual information are the same, one can define a measure for quantum correlations, namely the quantum discord as
\begin{equation}
D(\rho^{ab})=I(\rho^{ab})-C(\rho^{ab}).
\end{equation}
Main challenge in the calculation of quantum discord is the evaluation of classical correlations, since it requires a complex optimization over all measurements on the system. The reason that there is no general analytical results on quantum discord except for very few special cases, is due to this difficulty. It is important to note that quantum discord is dependent on which subsystem the measurements are done. Since making the measurements on spin-$j$ subsystem will make the optimization procedure even harder, in this work, all measurements are made on the spin-$1/2$ subsystem. Furthermore, QD can increase or decrease under local operations and classical communication (LOCC) if the LOCC is performed on the measured part of the system \cite{34,35,36,37}. This is a rather peculiar behavior since invariance under LOCC is the defining property of entanglement.

\section{Results and Discussion}
The bipartite state under consideration is composed of a spin-$j$ and a spin-$1/2$ subsystems. $SU(2)$ invariant states, are parameterized by a single parameter which will be denoted by $F$ throughout this paper. Density matrix for our system in total spin basis is given as \cite{25}
\begin{eqnarray}
\rho^{ab} &=\frac{F}{2j}\sum_{m=-j+1/2}^{j-1/2}|j-1/2, m\rangle\langle j-1/2, m| \\ \nonumber
& +\frac{1-F}{2(j+1)}\sum_{m=-j-1/2}^{j+1/2}|j+1/2, m\rangle\langle j+1/2, m|. \nonumber
\end{eqnarray}
We shall start by calculating the quantum mutual information. Bipartite density matrix has two eigenvalues $\lambda_1=F/2j$ and $\lambda_2=(1-F)/(2j+2)$ with degeneracies $2j$ and $2j+2$, respectively. On the other hand, the reduced density matrices of the subsystems can be found as $\rho^a=I_{2j+1}/(2j+1)$ and $\rho^b=I_{2}/2$ where $I_{2j+1}$ and $I_{2}$ is the identity matrix in the dimension of the Hilbert space for spin-$j$ and spin-$1/2$ particle, respectively. Note that both $\rho^a$ and $\rho^b$ are maximally mixed independent of $j$. Thus the mutual information of our system is
\begin{eqnarray}
I(\rho)&=S(\rho^a)+S(\rho^b)-S(\rho^{ab}) \\
&=1+\log_2(2j+1)+F\log_2\frac{F}{2j}+(1-F)\log_2\frac{1-F}{2j+2}. \nonumber
\end{eqnarray}

We now turn our attention to the calculation of the classical correlations. We will perform projective measurements on the spin-$1/2$ part of the density matrix. In order to do that, first we need to write the density matrix in the product basis. By using the Clebsh-Gordan coefficients for coupling a spin-$j$ to spin-$1/2$, density matrix in product basis can be written as
\begin{eqnarray}
\rho^{ab}= \frac{F}{2j}\sum_{m=-j+1/2}^{j-1/2} &a_-^2|m-1/2\rangle\langle m-1/2|\otimes |1/2\rangle\langle 1/2| \\ \nonumber
 &+a_-b_-(|m-1/2\rangle\langle m+1/2|\otimes |1/2\rangle\langle -1/2| \\ \nonumber
 &+|m+1/2\rangle\langle m-1/2|\otimes |-1/2\rangle\langle 1/2|) \\ \nonumber
 &+b_-^2|m+1/2\rangle\langle m+1/2|\otimes |-1/2\rangle\langle -1/2| \\ \nonumber
 +\frac{1-F}{2(j+1)}\sum_{m=-j-1/2}^{j+1/2} &a_+^2|m-1/2\rangle\langle m-1/2|\otimes |1/2\rangle\langle 1/2| \\ \nonumber
 &+a_+b_+(|m-1/2\rangle\langle m+1/2|\otimes |1/2\rangle\langle -1/2| \\ \nonumber
 &+|m+1/2\rangle\langle m-1/2|\otimes |-1/2\rangle\langle 1/2|) \\ \nonumber
 &+b_+^2|m+1/2\rangle\langle m+1/2|\otimes |-1/2\rangle\langle -1/2|. \nonumber
\end{eqnarray}
Here $a_{\pm}=\pm\sqrt{(j+1/2\pm m)/(2j+1)}$ and $b_{\pm}=\sqrt{(j+1/2\mp m)/(2j+1)}$ are the appropriate Clebsh-Gordon coefficients. Following \cite{5}, we can write any von Neumann measurement on $\rho^b$ as
\begin{equation}
\{B_k=V\Pi_kV^{\dag}:k=0,1\},
\end{equation}
where $\{\Pi_k=|k\rangle\langle k|:k=0,1\}$ and $V=tI+i\vec{y}\cdot\vec{\sigma}$, any unitary matrix in SU(2). Here, both $t$ and $\vec{y}$ are real and $t^2+y_1^2+y_2^2+y_3^2=1$. After the measurements are performed, $\rho^{ab}$ will transform into an ensemble of post-measurement states with their corresponding probabilities $\{\rho_k, p_k\}$. In order to calculate possible post-measurement states $\rho_k$ and their corresponding probabilities $p_k$, we write
\begin{eqnarray}
p_k\rho_k &=(I\otimes B_k)\rho^{ab}(I\otimes B_k)=(I\otimes V\Pi_kV^{\dag})\rho^{ab}(I\otimes V\Pi_kV^{\dag}) \\ \nonumber
&=(I\otimes V)(I\otimes \Pi_k)(I\otimes V^{\dag})\rho^{ab}(I\otimes V)(I\otimes \Pi_k)(I\otimes V^{\dag}). \nonumber
\end{eqnarray}
Since transformation of the usual Pauli matrices under $V$ and $\Pi_k$ is known \cite{5}, it is easier to calculate the post-measurement states when the spin-$1/2$ part of the density matrix is written in terms of them. In order to do that, we will use following identities
\begin{eqnarray}
|1/2\rangle\langle1/2|&=\frac{1}{2}[I+\sigma_3] \\ \nonumber
|1/2\rangle\langle -1/2|&=\frac{1}{2}[\sigma_1+i\sigma_2] \\ \nonumber
|-1/2\rangle\langle1/2|&=\frac{1}{2}[\sigma_1-i\sigma_2] \\ \nonumber
|-1/2\rangle\langle -1/2|&=\frac{1}{2}[I-\sigma_3]. \nonumber
\end{eqnarray}
We are now ready to use the transformation properties of Pauli matrices as given in \cite{5}
\begin{equation}
V^{\dag}\sigma_1V=(t^2+y_1^2-y_2^2-y_3^2)\sigma_1+2(ty_3+y_1y_2)\sigma_2+2(-ty_2+y_1y_3)\sigma_3,
\end{equation}
\begin{equation}
V^{\dag}\sigma_2V=2(-ty_3+y_1y_2)\sigma_1+(t^2+y_2^2-y_1^2-y_3^2)\sigma_2+2(-ty_1+y_2y_3)\sigma_3,
\end{equation}
\begin{equation}
V^{\dag}\sigma_3V=2(ty_2+y_1y_3)\sigma_1+2(-ty_1+y_2y_3)\sigma_2+(t^2+y_3^2-y_1^2-y_2^2)\sigma_3,
\end{equation}
and $\Pi_0\sigma_3\Pi_0=\Pi_0$, $\Pi_1\sigma_3\Pi_1=-\Pi_1$, $\Pi_j\sigma_k\Pi_j=0$ for $j=0,1$, $k=1,2$. We have calculated the probabilities of obtaining two possible post-measurement states as $p_0=p_1=1/2$ and the corresponding post-measurement states themselves as
\begin{eqnarray}
\rho_0= &\Bigg\{\sum_{m=-j}^{j} \left[\frac{1}{2j+1}-z_3\frac{m(2Fj+F-j)}{j(j+1)(2j+1)}\right]|m\rangle\langle m| \\ \nonumber
 &-(z_1+iz_2)\frac{\sqrt{j(j+1)-m(m+1)}(2Fj+F-j)}{2j(j+1)(2j+1)}|m\rangle\langle m+1|  \\ \nonumber
 &-(z_1-iz_2)\frac{\sqrt{j(j+1)-m(m+1)}(2Fj+F-j)}{2j(j+1)(2j+1)}|m+1\rangle\langle m|\Bigg\} \otimes V\Pi_0V^{\dag}
\end{eqnarray}
and
\begin{eqnarray}
\rho_1= &\Biggl\{\sum_{m=-j}^{j} \left[\frac{1}{2j+1}+z_3\frac{m(2Fj+F-j)}{j(j+1)(2j+1)}\right]|m\rangle\langle m| \\ \nonumber
 &+(z_1+iz_2)\frac{\sqrt{j(j+1)-m(m+1)}(2Fj+F-j)}{2j(j+1)(2j+1)}|m\rangle\langle m+1|  \\ \nonumber
 &+(z_1-iz_2)\frac{\sqrt{j(j+1)-m(m+1)}(2Fj+F-j)}{2j(j+1)(2j+1)}|m+1\rangle\langle m|\Biggr\} \otimes V\Pi_1V^{\dag}, \nonumber
\end{eqnarray}
where $z_1=2(-ty_2+y_1y_3)$, $z_2=2(ty_1+y_2y_3)$, $z_3=t^2+y_3^2-y_1^2-y_2^2$ with $z_1^2+z_2^2+z_3^2=1$.
The eigenvalues of the post-measurement states are the same and by inspection, they can be found as
\begin{equation}
\lambda_n^{\pm}=\frac{1}{2j+1}\pm\frac{j-n}{j(j+1)(2j+1)}|(F(2j+1)-j)|,
\end{equation}
where $n=0,\cdots, \lfloor j\rfloor$ for half-integer $j$ with $\lfloor .\rfloor$ being the floor function and $n=0,\cdots, j$ for integer $j$.

In calculation of the post measurement states, we have followed the way introduced in \cite{5}. Considering the symmetry of the states considered in this work, an alternative and a more direct way to obtain the eigenvalues of the post measurement states is present. Continuing directly from (8)
\begin{eqnarray}
p_k\rho_k &=(I\otimes V\Pi_kV^{\dag})\rho^{ab}(I\otimes V\Pi_kV^{\dag}) \\ \nonumber
&=(I\otimes V\Pi_kV^{\dag}(V\otimes V)\rho^{ab}(V^{\dag}\otimes V^{\dag})(I\otimes V\Pi_kV^{\dag}) \\ \nonumber
&=(I\otimes V\Pi_k)(V\otimes I)\rho^{ab}(V^{\dag}\otimes I)(I\otimes \Pi_kV^{\dag}) \\ \nonumber
&=(V\otimes V\Pi_k)\rho^{ab}(V^{\dag}\otimes\Pi_kV^{\dag}) \\ \nonumber
&=(V\otimes V)(I\otimes\Pi_k)\rho^{ab}(I\otimes\Pi_k)(V^{\dag}\otimes V^{\dag}). \\ \nonumber
\end{eqnarray}
Applying the projection operators to the spin-$1/2$ part of the density matrix, one can get the post measurement states as
\begin{equation}
p_0\rho_0=\frac{F}{2j}\sum_{m=-j+1/2}^{j-1/2}a_-^2|m-1\rangle\langle m-1|+\frac{1-F}{2(j+1)}\sum_{m=-j-1/2}^{j+1/2}a_+^2|m-1\rangle\langle m-1|
\end{equation}
and
\begin{equation}
p_1\rho_1=\frac{F}{2j}\sum_{m=-j+1/2}^{j-1/2}b_-^2|m-1\rangle\langle m-1|+\frac{1-F}{2(j+1)}\sum_{m=-j-1/2}^{j+1/2}b_+^2|m-1\rangle\langle m-1|.
\end{equation}
Since both of these matrices are diagonal and free of measurement parameters, it is straightforward to calculate the eigenvalues and eventually,
the QD of these states. The eigenvalues obtained from these post measurement states are equivalent to the ones presented in (15).

It can be clearly seen that that the eigenvalues do not depend on the measurement parameters. Therefore, calculation of the classical correlations do not require any optimization over the projective measurements. Then, the classical correlations can be written as
\begin{equation}
C(\rho^{ab})=S(\rho^a)-\sum_kp_kS(\rho_k^a)=\log_2(2j+1)+\sum_{n=0}^j\lambda_n^{\pm}\log_2(\lambda_n^{\pm}).
\end{equation}
Combining the above equation with (5), we have obtained an analytical expression for QD in the system under consideration
\begin{equation}
D(\rho^{ab})=1+F\log_2\frac{F}{2j}+(1-F)\log_2\frac{1-F}{2j+2}-\sum_{n=0}\lambda_n^{\pm}\log_2(\lambda_n^{\pm}),
\end{equation}
where $\lambda_n^{\pm}$ is given at (15).
\begin{figure}[H]
\begin{center}
\includegraphics[scale=0.5]{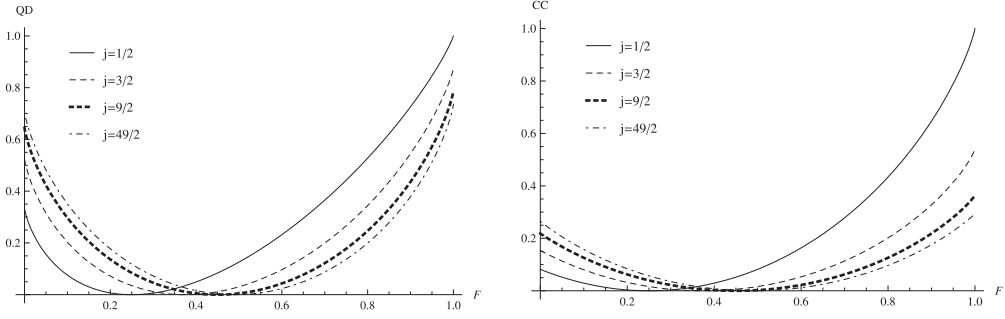}
\caption{On the left panel QD vs. F and on the right panel CC vs. F for $j=1/2$ $(d=2)$, $j=3/2$ $(d=4)$, $j=9/2$ $(d=10)$ and $j=49/2$ $(d=50)$.}
\end{center}
\end{figure}
In Fig. 1, we present our results on QD and $C(\rho^{ab})$ as a function of our system parameter $F$ for different dimensions. We recover the results obtained in \cite{5,38} in the special case of two spin-$1/2$ system. We know that for $\rho^{ab}$, the boundary between separable and entangled states is at $F_s=2j/(2j+1)$ \cite{25}, which is half of the value that both QD and $C(\rho^{ab})$ vanish $F_d=j/(2j+1)$. One can observe that as the dimension of the system increases, both QD and $C(\rho^{ab})$ increase in the region $F<F_d$ and decrease in the region $F>F_d$. Eventually, in the infinite $j$ limit, both of them become symmetric around the point $F=1/2$ where they are exactly zero. The symmetry around $F=1/2$ clearly starts to manifest itself at system dimensions as low as $j=9/2$ ($d=10$). The maximum value of QD is attained for $F=1$ for all system dimensions which corresponds to the state that is the projector on to the spin-$(j-1/2)$ subspace. It is important to note that as $j\rightarrow\infty$, our system becomes completely separable while QD remains finite except for a certain point, with its maximum value following a decreasing trend. This behavior can also be seen explicitly if we look at the large $j$ limit of (20) as
\begin{equation}
D(\rho^{ab})=1+F\log_2F+(1-F)\log_2(1-F)-\log_2(2j+1)-\sum_{n=0}^{j}\Lambda_n^{\pm}\log_2\Lambda_n^{\pm},
\end{equation}
where $\Lambda_n^{\pm}=1/2j\pm (j-n)|(2F-1)|/(2j^2)$. The symmetry point $F=1/2$ is apparent in the above equation and decreasing trend of the maximum value of QD can also be seen analytically as a function of $j$. In the same limit for $d\otimes d$ Werner states $F_s=F_d=1/2$ and QD is again symmetric around this point. Therefore, for $QD<1$, it is possible to find an entangled and a separable state possessing same amount of QD \cite{11}. From the right panel of Fig. 1, it is clear that classical correlations decay in the limit $j\rightarrow\infty$. However, its maximum settles to a fairly high value as compared to $d\otimes d$ Werner states.

\begin{figure}[H]
\begin{center}
\includegraphics[scale=0.5]{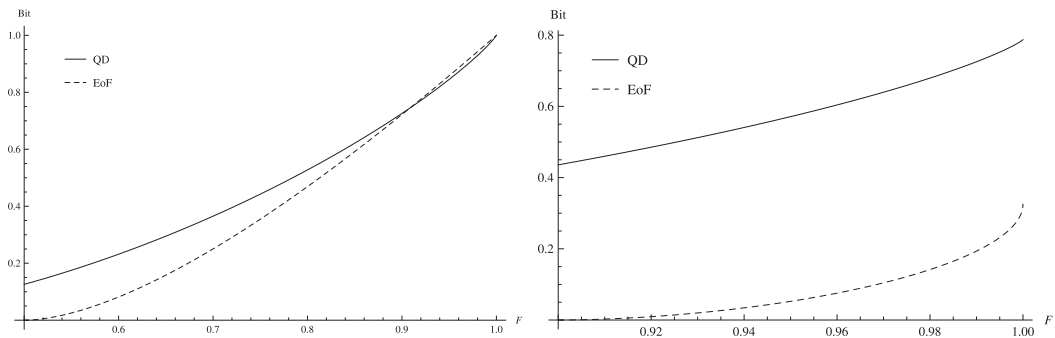}
\caption{QD (solid line) and EoF (dashed line) vs. $F$ for $j=1/2$ ($d=2$) (left panel) and for $j=9/2$ ($d=10$) (right panel)}
\end{center}
\end{figure}
We will now compare the amount of QD and entanglement possessed in our system. EoF for a spin-$1/2$ and a spin-$j$ $SU(2)$ invariant states is given by \cite{32}
\[ EoF= \left\{ \begin{array}{ll}
        0, & F\in[0,2j/(2j+1)] \\
        H\left(\frac{1}{2j+1}\left(\sqrt{F}-\sqrt{2j(1-F)}\right)^2 \right), & F\in[2j/(2j+1),1], \end{array} \right. \]
where $H(x)=-x\log x-(1-x)\log (1-x)$ is the binary entropy. In contrast to $d\otimes d$ Werner states, the point in the parameter space for which EoF becomes non-zero is dependent on $j$. In \cite{11}, it was shown that EoF becomes a general upper bound for QD in $d\otimes d$ Werner states. However, in figure 2, we can see that except $j=1/2$ case, QD always remains larger than EoF for all $F$ and the difference between these quantities increase as $j\rightarrow\infty$. Note that the region in which EoF remains zero covers the whole parameter space in the same limit.

\section{Conclusion}

We have analytically calculated the QD of a $SU(2)$ invariant system, consisting of a spin-$j$ and a spin-$1/2$ subsystems. We have compared our results with entanglement structure of these systems and QD of states having similar symmetries. It is known that a very small subset of the set of states addressed in this work possess entanglement as the dimension of the spin-$j$ particle becomes larger. We have shown that in the large $j$ limit, QD remains significantly larger than the entanglement. On the other hand, we have seen that maximum value of QD decreases with the increasing system size. Observation of $SU(2)$ invariant states in many real physical systems, make them a good candidate for utilization in quantum computing protocols that rely on QD.

\ack

The authors would like to thank G. Karpat for fruitful discussions, Yao-Kun Wang for an helpful comment, and an anonymous referee for important comments on the alternative way of finding the eigenvalues of the post measurement states. This work has been partially supported by the Scientiﬁc and Technological Research Council of Turkey (TUBITAK) under Grant No. 111T232.

\end{document}